\documentclass[aps,twocolumn,pra,superscriptaddress,showpacs,tightenlines]{revtex4-1}
\usepackage{amsmath}
\usepackage{graphicx}
\usepackage{color}
\usepackage{amsfonts}
\usepackage{txfonts}
\usepackage{lipsum} 
\usepackage{amsmath,epsfig,amssymb,subfigure,bm,dsfont}
\usepackage[colorlinks,citecolor=blue]{hyperref}

\begin{document}

\title{\textbf{Quantum signatures of transitions from stable fixed points to limit cycles in \\optomechanical systems}}
\author{Qing Xia Meng}
\affiliation{Department of Physics, College of Liberal Arts and Sciences, National University of Defense Technology, Changsha 410073, China}
\affiliation{Interdisciplinary Center for Quantum Information, National University of Defense Technology, Changsha 410073, China}
\author{Zhi Jiao Deng}
\email{dengzhijiao926@hotmail.com}
\affiliation{Department of Physics, College of Liberal Arts and Sciences, National University of Defense Technology, Changsha 410073, China}
\affiliation{Interdisciplinary Center for Quantum Information, National University of Defense Technology, Changsha 410073, China}
\author{Zhigang Zhu}
\affiliation{Department of Physics, Lanzhou University of Technology, Lanzhou, Gansu 730050, China}
\affiliation{School of Physical Science and Technology and Key Laboratory for Magnetism and Magnetic Materials of MOE, Lanzhou University, Lanzhou, Gansu 730000, China}
\author{Liang Huang}
\affiliation{School of Physical Science and Technology and Key Laboratory for Magnetism and Magnetic Materials of MOE, Lanzhou University, Lanzhou, Gansu 730000, China}

\begin{abstract}
Optomechanical systems, due to its inherent nonlinear optomechanical coupling, owns rich nonlinear dynamics of different types of motion. The interesting question is that whether there exist some common quantum features to infer the nonlinear dynamical transitions from one type to another. In this paper, we have studied the quantum signatures of transitions from stable fixed points to limit cycles in an optomechanical phonon laser system. Our calculations show that the entanglement of stable fixed points in the long run does not change with time, however, it will oscillate periodically with time at the mechanical vibration frequency for the limit cycles. Most strikingly, the entanglement quite close to the boundary line keeps as a constant, and it is very robust to the thermal phonon noise, as strong indications of this particular classical transitions.
\end{abstract}


\maketitle

\section{Introduction\label{sec:intro}}

Optomechanics, which deals with the nonlinear dynamics of coupled radiation
field and mechanical vibrations, has attracted huge recent attentions
\cite{1}. Ground state cooling is expected in many applications \cite{2},
therefore it is very important to make clear the quantum states and quantum
properties of the systems at low temperature. Due to the intrinsic nonlinear nature, the optomechanical system owns rich nonlinear dynamics such as bistability, limit cycle, and chaos \cite{3}. When the temperature goes down, the influence of quantum fluctuations becomes prominent and various quantum properties would also appear. An interesting question is that when the classical nonlinear dynamics change from one type to another, are there any signatures of these transitions in the corresponding quantum system?

There are already several related works \cite{4,5,6,7,8} in this regard. Reference \cite{4} shows that the time evolution of quantum entanglement is
periodic for limit cycles, while it exhibits beats-like behavior with two
distinct frequencies for quasiperiodic motion. And the most surprising feature is that the entanglement vanishes abruptly at the boundary of these two motions, as a strong quantum fingerprints of this particular transition. In a system of two coupled optomechanical cavities, the entanglement of two mechanical modes reveals a second-order phase transition type of change at the critical point from their in-phase to antiphase synchronization \cite{5}. Reference \cite{6} proposes new measures for quantum synchronization, and points out that their data are not sufficient to clarify the functional relationship between quantum synchronization and quantum discord. Another group also investigates the measure for quantum synchronization, and they find out that quantum discord behaves similarly to the measure of quantum synchronization based on their concrete optomechanical model \cite{7}. The entanglement in the bistable regime has also been analyzed, which will jump discontinuously along the hysteresis loop \cite{8}. Most of the previous works discuss only one set of parameters passing through the transition point. It is
natural to ask whether the changing quantum properties show common features no matter where to cross the boundary of two different types of nonlinear motions?

In this article, we will investigate the quantum signatures of transitions
from stable fixed points to limit cycles. Our discussions will be based on a two-dimensional phase diagram of an optomechanical phonon laser model
\cite{9}. The phonon laser, also referred to as mechanical self-sustained
oscillation \cite{1}, is essentially a limit cycle from the perspective of
nonlinear dynamics. It has been studied thoroughly \cite{10,11,12,13,14,15,16,17,18} and realized in recent experiments
\cite{9,19,20} in the context of optomechanics. The phonon laser in Ref.
\cite{9} is generated by the parametric down conversion process \cite{21}. The system will reach a stable fixed point in the long run when the driving power is not very strong, but undergoes a limit cycle motion once above certain driving threshold. Our aim is to look for the common changing features of quantum entanglement around the boundaries of these two nonlinear motions. To do that, we choose several different paths to cross the boundaries. Our calculations show that the entanglement for the stable fixed points does not change with time, while it oscillates at the mechanical frequency for the limit cycles. The most striking phenomenon is that the entanglement of those points very close to their boundary line is a constant, and it is very robust to the mechanical thermal noise, as obvious quantum signatures of this nonlinear dynamical transitions from one to another. Our paper is organized as follows, in Sec. II, we will introduce the physical model and derive its equations of motion. In Sec. III, we first present the classical equations of motion and give a two-dimensional phase diagram regarding the strength and detuning of the driving laser, and then discuss the classical nonlinear
dynamics along three different paths in the phase diagram. In Sec. IV, we show the general procedures to calculate the quantum entanglement, and study how it will change along the above mentioned three paths. In Sec. V, we summarize our results.
\section{PHYSICAL SYSTEM \label{sec:model}}

The optomechanical system in the experiment of Ref. \cite{9} consists of two
coupled cavity modes, one of which is coupled to a mechanical mode by the
radiation pressure force, and the other is driven by an input laser. The
Hamiltonian of the whole system is,%
\begin{eqnarray}
\hat{H}&=&\hbar\omega_{a}(\hat{a}_{1}^{\dagger}\hat{a}_{1}+\hat{a}_{2}^{\dagger
}\hat{a}_{2})+\hbar J(\hat{a}_{1}^{\dagger}\hat{a}_{2}+\hat{a}_{1}\hat{a}%
_{2}^{\dagger})-\hbar g\hat{a}_{2}^{\dagger}\hat{a}_{2}\hat{q}\nonumber\\
&&+\frac
{\hbar\omega_{m}}{2}(\hat{p}^{2}+\hat{q}^{2})+i\hbar\Lambda(\hat{a}%
_{1}^{\dagger}e^{-i\omega_{L}t}-\hat{a}_{1}e^{i\omega_{L}t}).
\end{eqnarray}
where the two localized cavity modes have the same frequency $\omega_{a}$, and their tunnelling rate is denoted by $J$. The mechanical mode with frequency $\omega_{m}$ is coupled to cavity mode 2 by a constant coupling strength $g$. The operators $\hat{q}=\frac{1}{\sqrt{2}}(\hat{b}^{\dagger}+\hat{b})$,
$\hat{p}=\frac{1}{\sqrt{2}i}(\hat{b}-\hat{b}^{\dagger})$ represent the
dimensionless position and momentum of the mechanical mode respectively. The last term describes the driving of cavity mode 1 by a laser with frequency $\omega_{L}$ and amplitude $\Lambda$.

A proper analysis of the system must include photon losses in the cavity and the Brownian noise acting on the mechanical vibration. This can be
accomplished by considering the following set of nonlinear Langevin equations (written in the interaction picture with respect to $\hbar\omega_{L}(\hat
{a}_{1}^{\dagger}\hat{a}_{1}+\hat{a}_{2}^{\dagger}\hat{a}_{2})$) \cite{22},%
\begin{align}
\dot{\hat{a}}_{1} &  =(i\Delta-\frac{\kappa}{2})\hat{a}_{1}%
-iJ\hat{a}_{2}+\Lambda+\sqrt{\kappa}\hat{a}_{in,1},\\
\dot{\hat{a}}_{2} &  =(i\Delta-\frac{\kappa}{2})\hat{a}_{2}%
-iJ\hat{a}_{1}+ig\hat{a}_{2}\hat{q}+\sqrt{\kappa}\hat{a}_{in,2},\\
\dot{\hat{q}} &  =\omega_{m}\hat{p},\\
\dot{\hat{p}} &  =g\hat{a}_{2}^{\dagger}\hat{a}_{2}-\omega_{m}%
\hat{q}-\gamma_{m}\hat{p}+\hat{\xi}.
\end{align}
Here $\Delta=\omega_{L}-\omega_{a}$denotes the laser detuning from the
cavity resonance, $\gamma_{m}$ is the mechanical damping rate, and $\kappa$ is
the optical intensity decay rate. The operators $\hat{a}_{in,1}$, $\hat
{a}_{in,2}$ are the vacuum radiation input noise. Their mean values satisfy
$\langle\hat{a}_{in,j}(t)\rangle=0$, and their only nonzero correlation
functions fulfill $\langle\hat{a}_{in,j}(t)\hat{a}_{in,j^{\prime}}^{\dagger
}(t^{\prime})\rangle=\delta_{jj^{\prime}}\delta(t-t^{\prime})$ with $j=1,2$. The Hermitian Brownian noise operator $\hat{\xi}$ with zero mean value, satisfies a delta-correlated function $\frac{1}{2}\left\langle \hat{\xi}(t)\hat{\xi}(t^{^{\prime}})+\hat{\xi}(t^{^{\prime}})\hat{\xi}(t)\right\rangle =\gamma_{m}(2\overline{n}+1)\delta(t-t^{^{\prime}})$ in the limit of high
mechanical quality factor \cite{4,7,23}, i.e., $Q=\omega_{m}/\gamma_{m}\gg1$,
where $\overline{n}=(\exp(\frac{\hbar\omega_{m}}{k_{B}T})-1)^{-1}$ is the mean
thermal phonon number at temperature $T$, and $k_{B}$ is Boltzmann's constant.

The mechanism to generate the phonon laser can be understood more clearly if
we transform to the basis with supermodes defined as $\hat{c}_{1}=\frac
{1}{\sqrt{2}}(\hat{a}_{1}+\hat{a}_{2})$, $\hat{c}_{2}=\frac{1}{\sqrt{2}}%
(\hat{a}_{1}-\hat{a}_{2})$. The Langevin equations are now in the following
forms,%
\begin{align}
\dot{\hat{c}}_{1}  &  =(i(\Delta-J)-\frac{\kappa}{2})\hat{c}%
_{1}+\frac{ig}{2}(\hat{c}_{1}-\hat{c}_{2})\hat{q}+\frac{\Lambda}{\sqrt{2}%
}+\sqrt{\kappa}\hat{c}_{in,1},\\
\dot{\hat{c}}_{2}  &  =(i(\Delta+J)-\frac{\kappa}{2})\hat{c}%
_{2}-\frac{ig}{2}(\hat{c}_{1}-\hat{c}_{2})\hat{q}+\frac{\Lambda}{\sqrt{2}%
}+\sqrt{\kappa}\hat{c}_{in,2},\\
\dot{\hat{q}}  &  =\omega_{m}\hat{p},\\
\dot{\hat{p}}  &  =\frac{g}{2}(\hat{c}_{1}^{\dagger}\hat{c}%
_{1}+\hat{c}_{2}^{\dagger}\hat{c}_{2}-\hat{c}_{1}^{\dagger}\hat{c}_{2}-\hat
{c}_{2}^{\dagger}\hat{c}_{1})-\omega_{m}\hat{q}-\gamma_{m}\hat{p}+\hat{\xi
}.
\end{align}
where $\hat{c}_{in,1}=\frac{1}{\sqrt{2}}(\hat{a}_{in,1}+\hat{a}_{in,2})$,
$\hat{c}_{in,2}=\frac{1}{\sqrt{2}}(\hat{a}_{in,1}-\hat{a}_{in,2})$, obeying
similar correlation functions as for $\hat{a}_{in,1}$ and $\hat{a}_{in,2}$ .
The eigenfrequencies for $c_{1}$ and $c_{2}$ modes in the interaction picture
are $-(\Delta-J)$, $-(\Delta+J)$ respectively. If their frequency difference
$2J$ is near resonant with the mechanical frequency $\omega_{m}$, i.e.,
$2J\simeq\omega_{m}$, then an efficient driving of $c_{1}$ mode with
$\Delta\simeq J$ could lead to a parametric down conversion process via the
interaction term $\hat{c}_{1}^{\dagger}\hat{c}_{2}\hat{b}+\hat{c}_{2}%
^{\dagger}\hat{c}_{1}\hat{b}^{\dagger}$, which means that when one photon in
$c_{1}$ mode disappear, meanwhile a photon in $c_{2}$ mode and a phonon are
born. When the driving is above the threshold power, coherent oscillation
(i.e., mechanical lasing) would occur in the mechanical mode. Moreover, this
two-mode squeezing interaction term will inevitably result in the quantum
entanglement between the optical $c_{2}$ mode and the mechanical mode, as
discussed in many previous works \cite{24,25,26,27,28,29,30,31}%
.

\section{NONLINEAR DYNAMICS}

The Eq. (3) in the regime of weak coupling $g\ll\kappa$ and moderate driving
$\Lambda$ can be solved by the mean-field approximation \cite{1}, in which
quantum operators are separated into $\hat{O}=\langle\hat{O}\rangle+\delta
\hat{O}$, where $\langle\hat{O}\rangle\equiv O$ is the mean field describing
the classical behavior of the system, and $\delta\hat{O}$ is the quantum
fluctuation with zero mean value around the classical orbit. In this section,
we will focus on the classical dynamics of the system. The equations of motion
for the classical mean fields form a set of nonlinear differential equations
given by%
\begin{align}
\dot{c}_{1}  &  =[i(\Delta-J)-\frac{\kappa}{2}]c_{1}+\frac{ig}%
{2}(c_{1}-c_{2})q+\frac{\Lambda}{\sqrt{2}},\\
\dot{c}_{2}  &  =[i(\Delta+J)-\frac{\kappa}{2}]c_{2}-\frac{ig}%
{2}(c_{1}-c_{2})q+\frac{\Lambda}{\sqrt{2}},\\
\dot{q}  &  =\omega_{m}p,\\
\dot{p}  &  =-\omega_{m}q-\gamma_{m}p+\frac{1}{2}g(c_{1}^{\ast
}c_{1}+c_{2}^{\ast}c_{2}-c_{1}^{\ast}c_{2}-c_{2}^{\ast}c_{1}).
\end{align}
which is obtained by averaging on both sides of Eq. (3), and approximates
$\langle\hat{F}\hat{G}\rangle$ with $\langle\hat{F}\rangle\langle\hat
{G}\rangle$.

First, we do the stability analysis of the fixed points \cite{3} in Eq. (4). The fixed points are the solutions after letting all the first-order
derivatives $\dot{O}$ to be 0. Their stability can be judged by the linearized Langevin equations for the quantum fluctuation operators, which can be expressed in the compact matrix form as \cite{8,32},%
\begin{equation}
\dot{u}(t)=S(t)u(t)+n(t).
\end{equation}
where we have defined $u^{T}(t)$=$(\delta\hat{X}_{1}(t),\delta\hat{Y}_{1}(t),\delta\hat{X}_{2}(t),\delta\hat{Y}_{2}(t),\\
\delta\hat{q}(t),\delta\hat{p}(t))$ and the input noise operators $n^{T}(t)$=$(\sqrt{\kappa}\hat {X}_{in,1}(t),\sqrt{\kappa}\hat{Y}_{in,1}(t),\sqrt{\kappa}\hat{X}_{in,2}(t),\sqrt{\kappa}\hat{Y}_{in,2}(t),0,\hat{\xi}(t))$, with quadrature operators $\delta\hat{X}_{j}$=$\frac{1}{\sqrt{2}}(\delta\hat{c}_{j}$+$\delta\hat{c}_{j}^{\dagger})$, $\delta\hat{Y}_{j}$=$\frac{1}{\sqrt{2}i}(\delta\hat
{c}_{j}-\delta\hat{c}_{j}^{\dagger})$, and the corresponding Hermitian input noise operators $\hat{X}_{in,j}=\frac{1}{\sqrt{2}}(\hat{c}_{in,j}+\hat{c}_{in,j}^{\dagger})$, $\hat{Y}_{in,j}=\frac{1}{\sqrt{2}i}(\hat{c}%
_{in,j}-\hat{c}_{in,j}^{\dagger})$ $(j=1,2)$. Furthermore, the coefficient matrix $S$ has the form,
\begin{widetext} 
	\begin{eqnarray} 
	S(t)=%
	\begin{pmatrix}
	-\frac{\kappa}{2} & -(\Delta-J)-\frac{g}{2}q & 0 & \frac{g}{2}q & -\frac{g}%
	{2}(y_{1}-y_{2}) & 0\\
	(\Delta-J)+\frac{g}{2}q & -\frac{\kappa}{2} & -\frac{g}{2}q & 0 & \frac{g}%
	{2}(x_{1}-x_{2}) & 0\\
	0 & \frac{g}{2}q & -\frac{\kappa}{2} & -(\Delta+J)-\frac{g}{2}q & \frac{g}%
	{2}(y_{1}-y_{2}) & 0\\
	-\frac{g}{2}q & 0 & (\Delta+J)+\frac{g}{2}q & -\frac{\kappa}{2} & -\frac{g}%
	{2}(x_{1}-x_{2}) & 0\\
	0 & 0 & 0 & 0 & 0 & \omega_{m}\\
	g(x_{1}-x_{2}) & g(y_{1}-y_{2}) & -g(x_{1}-x_{2}) & -g(y_{1}-y_{2}) &
	-\omega_{m} & -\gamma_{m}%
	\end{pmatrix}.
	\end{eqnarray} 
\end{widetext}

Here $x_{j}$, $y_{j}$ are the real part and imaginary part of the complex
amplitude $c_{j}$ $(j=1,2)$ respectively. The dynamics of matrix $S$ depends
on the time evolution of Eq. (3) under the assumption that the quantum
fluctuations always follow the classical orbit, which is guaranteed as long as
none of the Lyapunov exponents in the corresponding classical equations is
positive \cite{4}. For analysis of the stability, the linearization is
performed around the fixed point. The system is stable only if all eigenvalues
of matrix $S$ evaluated at the fixed point have negative real parts.

\begin{figure}[tbp]
\center
\includegraphics[width=0.47 \textwidth]{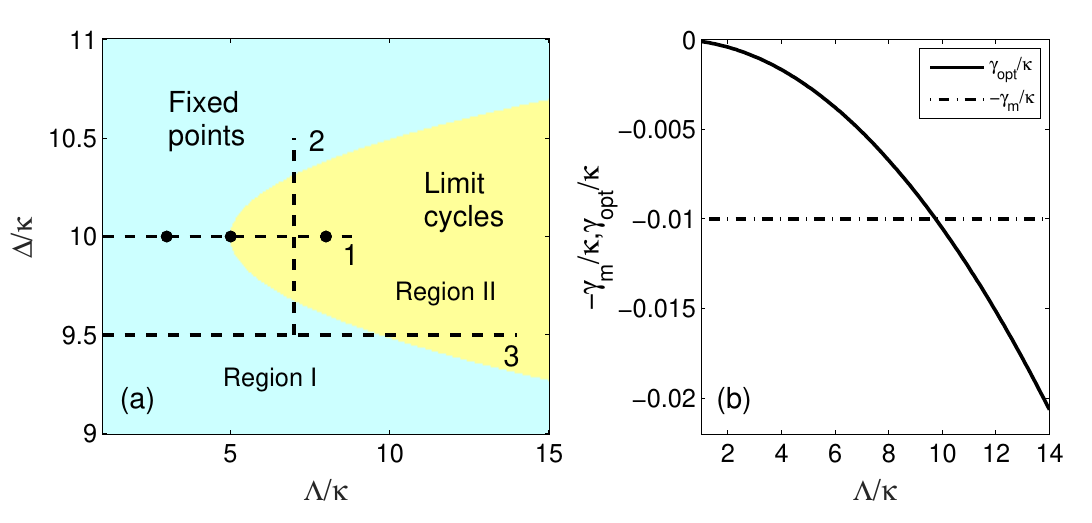}
\caption{(a) Phase diagram describing the long time dynamical behaviours of the optomechanical phonon laser with the parameters $J/\kappa=10$, $\omega_{m}/\kappa=20$, $g/\kappa=0.02$, $\gamma_{m}/\kappa=0.01$. The three dashed lines labeled with 1, 2, 3 denotes path 1, path 2 and path 3 respectively (path 1: resonant driving of $c_{1}$ mode with $\Delta=J$; path 2: going vertically in the diagram with $\Lambda/\kappa=7$; path 3: driving $c_{1}$ mode with detuning $\Delta/\kappa=9.5$). The time evolution of the three marked points on path 1 from left to right with $\Lambda/\kappa=3,5.01,8$, will be shown below. (b) Calculation of the lasing threshold value (or transition point) for path 3 by finding the intersection point of the two lines $\gamma_{opt}(\Lambda)$ and $-\gamma_{m}$.}
\label{fig1}
\end{figure}
In Fig. 1(a), we choose $2J=\omega_{m}$ and plot the two-dimensional phase
diagram with respect to the driving strength $\Lambda$ and driving detuning
$\Delta$. The system will eventually arrive at the fixed points in region I,
by contrast it will settle into the limit cycles in region II. The mechanical
freedom in the latter case conducts an approximately sinusoidal oscillation at
its unperturbed frequency, i.e., $q(t)=q_{0}+A\cos(\omega_{m}t)$ with shifted
equilibrium position $q_{0}$ and amplitude $A$. The threshold value for lasing
can be obtained by demanding that the effective mechanical damping rate
$\gamma_{eff}=\gamma_{m}+\gamma_{opt}=0$ \cite{11}, where $\gamma_{opt}$ is
the optomechanical damping rate induced by the radiation pressure force. We
calculate the mechanical susceptibility \cite{1}, and get $\gamma_{opt}%
=\omega_{m}\left\vert \alpha_{2}\right\vert ^{2}g^{2}\frac{2\kappa
	\Delta(3B^{2}-2B(\omega_{m}^{2}+\Delta^{2})-(\omega_{m}^{2}-\Delta^{2}%
	)^{2}-B\kappa^{2})}{((B-(\omega_{m}+\Delta)^{2})^{2}+\kappa^{2}(\omega
	_{m}+\Delta)^{2})((B-(\omega_{m}-\Delta)^{2})^{2}+\kappa^{2}(\omega_{m}%
	-\Delta)^{2})}$, with $B=J^{2}+\frac{\kappa^{2}}{4}$, and $\alpha_{2}%
=(c_{1}-c_{2})/\sqrt{2}$ evaluated at the corresponding fixed point. 
As shown in Fig. 1(b), $\gamma_{opt}$ is negative and it decreases with the driving
amplitude. The intersection point of the two lines $\gamma_{opt}(\Lambda)$ and
$-\gamma_{m}$ indicates the driving threshold $\Lambda_{th}$. 
It becomes
larger when the driving detuning goes away from the resonant case $\Delta=J$.

\begin{figure}[ht]
	\center
	\includegraphics[width=0.47 \textwidth]{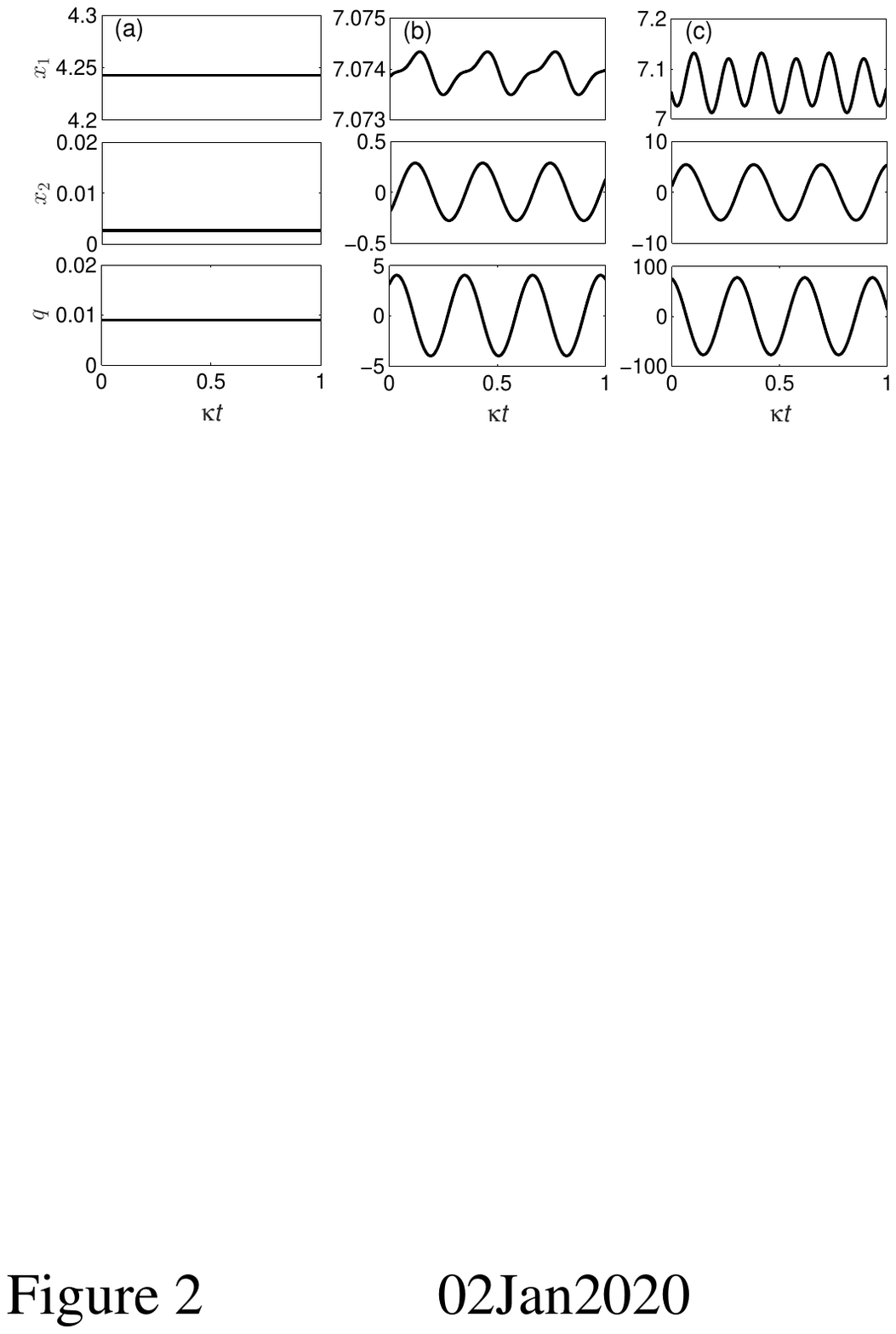}
	\caption{Long time dynamical behaviours for the three marked points from the left to right in Fig 1 (a) corresponds to the (a), (b), and (c) subplots here.}
	\label{fig2}
\end{figure}
To get more insights into the interplay between nonlinear dynamics and quantum
entanglement, we choose three typical paths to cross the boundaries (see Fig.
1(a)). The long time behavior of the three points marked on path 1 in Fig.
1(a) is explicitly displayed in Fig. 2. 
All the variables keep constant values
at the fixed point (the left point on path 1), while in region II (the middle
and right points on path 1) they oscillate with time at the mechanical
frequency $\omega_{m}$. The middle point described in Fig. 2(b) is very close
to the boundary, the oscillation for $x_{1}$ has only one maximum and one
minimum within one cycle. As we move away a little bit from the boundary, the
number of oscillation extrema for $x_{1}$ doubles (see Fig. 2(c)), developing
into the period-2 orbit \cite{33}. 
\begin{figure}[ht]
	\center
	\includegraphics[width=0.47 \textwidth]{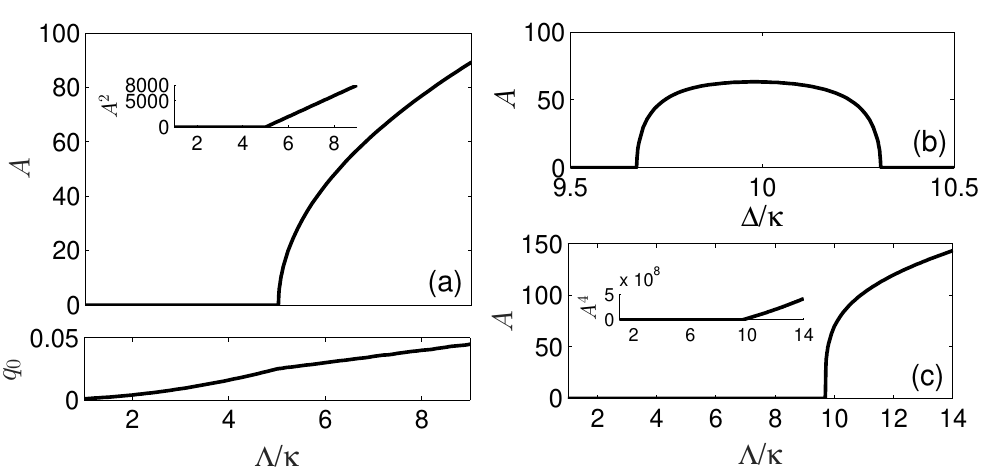}
	\caption{Mechanical oscillation amplitude $A$ on path 1, 2, 3 depicted in (a),
		(b), and (c) subplots respectively. The corresponding shifted equilibrium
		position $q_{0}$ on path 1 is also included in (a).}
	\label{fig3}
\end{figure}
In Fig. 3, we plot the mechanical
oscillation amplitude $A$ on the three paths. Path 1 in Fig. 3(a) represents
the resonant driving of $c_{1}$ mode, with the amplitude starts from $A=0$, is
an example of a Hopf bifurcation. The amplitude $A$ is proportional to
$\sqrt{\Lambda-\Lambda_{th}}$, and the bottom part shows the corresponding new
equilibrium position $q_{0}$ pushed by the radiation pressure force. The
stronger the input driving, the more the mechanical resonator will be shifted.
Until to some extent, it oscillates coherently. Path 2 in Fig. 3(b) goes
vertically in the phase diagram, and passes the boundary twice. The amplitude
near the boundary has a similar square root relationship as in path 1, i.e.,
$A\propto$ $\sqrt{|\Delta-\Delta_{th}|}$, where $\Delta_{th}$ is the detuning
at the boundary. Path 3 in Fig. 3(c) introduces some detuning in the driving
of $c_{1}$ mode. The amplitude $A$, which is proportional to $\sqrt[4]%
{\Lambda-\Lambda_{th}}$, increases more rapidly in the vicinity of the
threshold than the resonant driving.

\section{QUANTUM ENTANGLEMENT\label{sec:opensys}}

To check whether there exist quantum signatures of this classical transition,
we calculate the degree of quantum entanglement by using the logarithmic
negativity \cite{34}. The quantum statistical properties of the system can be
investigated through the small fluctuations of the operators around the
time-dependent mean values evolving according to Eq. (4). The standard
linearization \cite{35} around the classical orbit gives rise to Eq. (5).
Since the equations are linear, the fluctuations will remain Gaussian if the
input noises are Gaussian. In this case, the properties of quantum
fluctuations are fully characterized by the covariance matrix $V$, with its
elements defined by $V_{ij}=\frac{1}{2}(\langle u_{i}(t)u_{j}(t)+u_{j}%
(t)u_{i}(t)\rangle)$. The equation of motion for the covariance matrix is
governed by \cite{36},%
\begin{equation}
\dot{V}(t)=S(t)V(t)+V(t)S^{T}(t)+D.
\end{equation}
where $D=\text{diag}(\frac{\kappa}{2},\frac{\kappa}{2},\frac{\kappa}{2}%
,\frac{\kappa}{2},0,\gamma_{m}(2\overline{n}+1))$ is the diffusion matrix. The
optical $c_{2}$ mode and the mechanical mode are entangled, and their
entanglement is related to the covariance matrix $W$ between these two modes,
which is a submatrix of $V$,%
\begin{equation}
W=\left(
\begin{array}
[c]{cccc}%
V_{33} & V_{34} & V_{35} & V_{36}\\
V_{43} & V_{44} & V_{45} & V_{46}\\
V_{53} & V_{54} & V_{55} & V_{56}\\
V_{63} & V_{64} & V_{65} & V_{66}%
\end{array}
\right)  =\left(
\begin{array}
[c]{cc}%
M & C\\
C^{T} & N
\end{array}
\right)  .
\end{equation}
with $M$, $N$, $C$ being $2\times2$ matrices. $M$ and $N$ account for the
local properties of the $c_{2}$ mode and the mechanical mode, respectively,
while $C$ describes intermode correlations. The logarithmic negativity can be
obtained with the formula $E_{N}=\max[0,-\ln2\eta^{-}]$, where $\eta
^{-}=2^{-\frac{1}{2}}\{\sum(W)-[\sum(W)^{2}-4\det W]^{\frac{1}{2}}\}^{\frac
	{1}{2}}$, and $\sum(W)=\det(M)+\det(N)-2\det(C)$ \cite{35}. We are interested
in the long time behavior of the entanglement. In our numerical integration of
Eqs. (4) and (7), we start with a set of random initial values for $V$,
$c_{1}$, $c_{2}$, $q$, $p$ until $E_{N}$ reaches a steady state. The
entanglement in region I will evolve to a constant value, while in the region
II it oscillates periodically with the mechanical frequency. Since the quantum
fluctuations follow the classical orbit, it is not surprising that the
entanglement has similar time dependence as the classical dynamics, either
stationary or periodic. The linearization method to calculate the entanglement
for limit cycles has been used in several recent works \cite{4,5,7,37}. All of
them are discussed in the weak coupling and strong driving regime. In the
opposite case of strong coupling and weak driving regime, there are works that
have shown the phase diffusion phenomenon for limit cycles with full
simulation of master equations \cite{11,38}. In principle, the random noise
will make the steady state distribution smear out around the circle, in
contrast to the point-like picture assumed above. But since in our case the
optomechanical coupling is weak and the temperature considered is very low,
the influence of noise should be relatively small, leading to much longer
transient time before any phase diffusion significantly to happen. In such a
case, the point-like picture is still meaningful. The most strict way to check
is to do the full simulations of master equations, which is impossible in our
parameter regime due to the huge Hilbert space involved. This is an open
question and deserves further studies.
\begin{figure*}[ht]
	\center
	\includegraphics[ width=0.85 \textwidth]{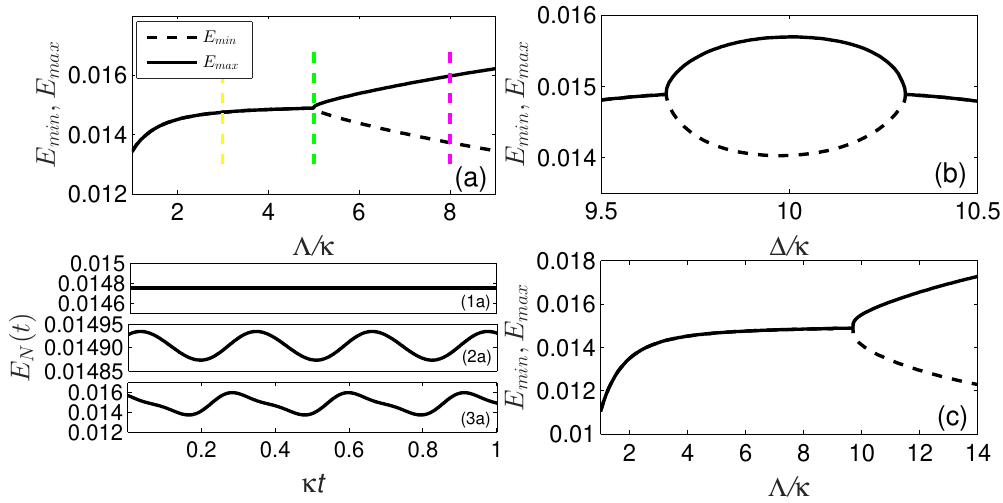}
	\caption{Steady state entanglement with $\overline{n}=0$ for path 1, 2, 3
		plotted in (a), (b), and (c) subplots respectively. The details of $E_{N}(t)$
		for the three marked points analyzed above, which also correspond to the three
		vertical dashed lines in (a) from left to right, are shown in (1a) (2a) (3a) respectively.}
	\label{fig4}
\end{figure*}

In Fig. 4, we plot the steady state entanglement of the three typical paths at
zero temperature. $E_{N}(t)$ varies over time within a certain range of
values, and we denote its maximum and minimum values as $E_{\max}$ and
$E_{\min}$ respectively. The entanglement for path 1 is depicted in Fig. 4(a),
where the two lines for $E_{\max}$ and $E_{\min}$ coincide below the
threshold, increase as approaching the transition point, where they start to
separate apart more and more with increasing driving amplitude. We give the
details of $E_{N}(t)$ for the three marked points (see Fig. 4 (1a) (2a) (3a)).
The entanglement for a stable fixed point is a constant and does not change
with time. Beyond the threshold, for the point that is close to the boundary,
the entanglement $E_{N}(t)$ oscillates in a symmetric sinusoidal form. As the
point moves away from the boundary, $E_{N}(t)$ gets tilted over time. This is
related to the emergence of period-2 orbit mentioned above. Figure 4(b) shows
the entanglement for path 2, which has two bifurcations corresponding to
passing the boundary twice and also demonstrates the tendency of increase
before the bifurcations. The maximum entanglement is achieved at some place in
between, where the mechanical oscillation amplitude $A$ is comparatively
large. The features for path 3 in Fig. 4(c) are quite similar, however, the
change at the bifurcation is much steeper, which is due to the rapid increase
of the amplitude $A$ near the threshold.
\begin{figure}[tbp]
	\center
	\includegraphics[width=0.47 \textwidth]{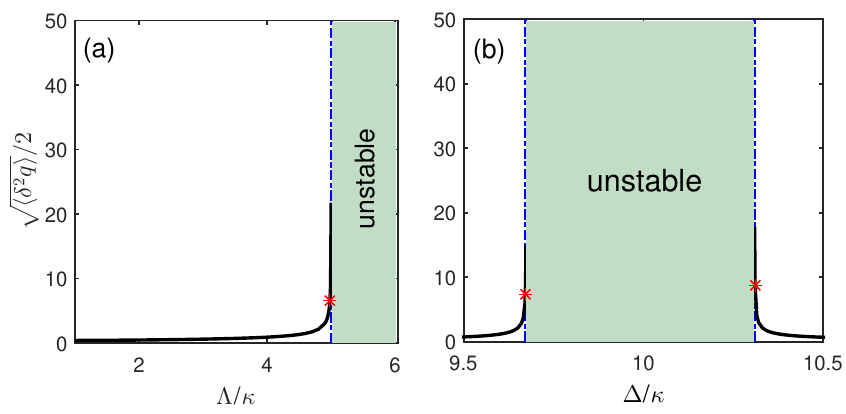}
	\caption{Mechanical fluctuation radius $\frac{1}{2}\sqrt{\langle\delta^{2}%
			\hat{q}\rangle}\left(  \simeq\frac{1}{2}\sqrt{\langle\delta^{2}\hat{p}\rangle
		}\right)  $ for stable fixed points of path 1 and path 2 shown in (a) and (b)
		respectively. Dashed lines are the boundary lines and red stars are the
		nearest points to the boundary chosen in our numerical calculations.}
	\label{fig5}
\end{figure}

The most interesting phenomenon is that the entanglement of those points quite
close to the boundary line is a constant and it is the maximum entanglement
for all the stable fixed points, which is a strong quantum fingerprint for the
transition from stable fixed points to limit cycles. Here, we emphasize that
the points can never be exactly on the boundary due to the numerical
discreteness, either on its left side or right side. As shown in Fig. 5, there
is a tendency of rapid increase of mechanical fluctuations in a very tiny
range approaching the boundary, which makes the linearization methods fail to
apply. So we exclude this tiny range in our calculations. However, the nearest
points to the boundary (see red stars in Fig. 5) we have chosen are good
enough to indicate the transition position. Note that although the parameter
values in region I of Fig. 1(a) are all for stable fixed points, the positions
of the fixed points in the parameter space are generally different.
Particularly, the four points closest to the boundary that the three paths in
Fig. 1(a) encounter have different positions, but they have the same
entanglement. We have also checked randomly many other points quite close to
the boundary, the entanglement keeps the same. For the parameters chosen in
Fig. 4, the constant is about 0.01488. We guess this should be related to the
function of boundary line, which is contained in the expression of the
entanglement, and leads to a constant value just quite near the boundary. But
the analytical calculation of this entanglement is too complicated for our model.
\begin{figure}[ht]
	\center
	\includegraphics[ width=0.47 \textwidth]{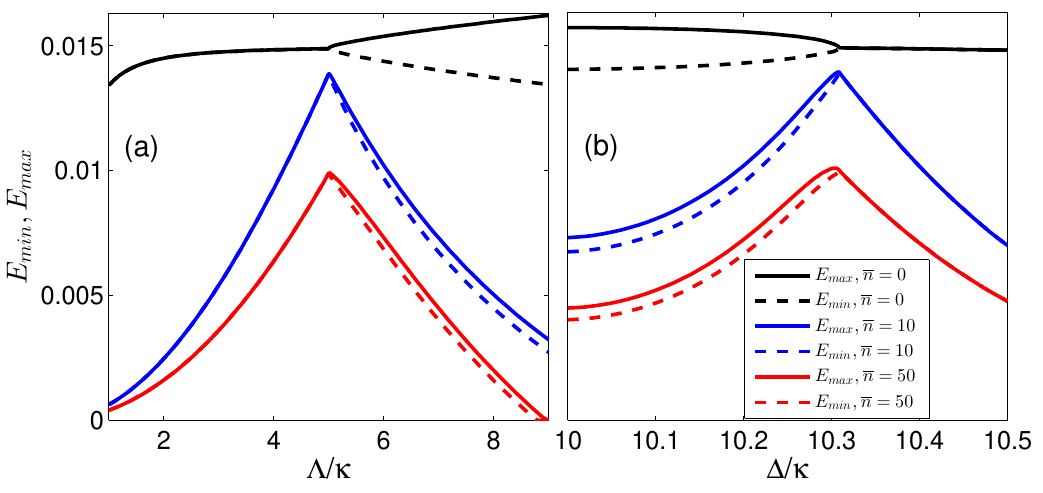}
	\caption{Temperature influences of the steady state entanglement for path 1 in
		(a) and path 2 in (b).}
	\label{fig6}
\end{figure}

We now consider the influence from the temperature. The entanglement of path 1
and path 2 with different mean thermal phonon numbers is given by Fig. 6. The
entanglement on both sides of the transition point falls down obviously with
the increase of temperature, while the entanglement quite near the boundary is
very robust to the presence of thermal mechanical noise. It decreases
relatively slower with the rising temperature, but keeps as a constant along
the boundary line. In the limit cycle region, the difference between the
maximum and minimum of entanglement, i.e., $E_{\max}-E_{\min}$, also decreases
with higher temperature. $E_{\min}$ touches zero first, and then $E_{\max}$
will follow, which means that there is no entanglement any more, for example,
the situation in Fig. 6(a) with $\Lambda/\kappa=9$ and $\overline{n}=50$.

\section{CONCLUSION\label{sec:conclusion}}

To summarize, we have studied how the quantum entanglement changes from stable
fixed points to limit cycles in an optomechanical phonon laser system, with
the aim of finding out the quantum signatures of this particular nonlinear
dynamical transitions. We pick out three different paths to cross the
boundary, and analyze their nonlinear dynamics and quantum entanglement
properties respectively. Our calculations show that indeed there are some
quantum features in common to indicate this classical transition: 1) The
quantum entanglement for the stable fixed points is a constant number, while
it oscillates with time at the mechanical frequency for the limit cycles. The
transition point is at which this oscillation starts to happen; 2) The
entanglement of the stable fixed points increases as approaching the
transition boundary, and reaches their maximum value quite close to the
boundary. Most strikingly, the entanglement of those points quite close to the
boundary line is a constant, which is a strong signal for the indication of
the transition border line; 3) The entanglement around the boundary line is
very robust to the influence of thermal noise, that it decreases relatively
slower with increasing temperature. Furthermore, even at finite temperature,
although the entanglement decreases, it has the same value along the boundary
line. Thus we can still easily find out the transition boundary by the amount
of quantum entanglement. In a word, we have investigated the fundamental
problem of quantum manifestations of transition between different types of
motions in nonlinear dynamical systems, which deserves much more efforts in
the future for transitions between other more complex dynamical behaviors.

\begin{acknowledgments}
Z. J. Deng is grateful to Florian Marquardt, Ying-Cheng Lai, Talitha Weiss and
Jie-Qiao Liao for useful discussions. This work is supported by the National
Natural Science Foundation of China under Grant Nos. 11574398, 11775101 and
61632021, the National Basic Research Program of China under Grant No.
2016YFA0301903, and by the Natural Science Foundation of Hunan Province of
China under Grant No. 2018JJ2467.
\end{acknowledgments}

\end{document}